\newcommand{\uvby}{{\em uvby} }
\newcommand{\uvbyb}{{\em uvby}$\beta$ }
\newcommand{\hb}{H$\beta$ }
\newcommand{\ha}{H$\alpha$ }

\newcommand{\cuno}{$c_{1}$ }

\newcommand{\byo}{$(b-y)_{0}$ } 
\newcommand{\mv}{$M_{V}$ }

\newcommand{\ra}{$t/t_{\rm MS}$ }
\newcommand{\sm}{M$_{\odot}$ }
\documentclass{aa}  

\begin{document}

\thesaurus{  05 
              (08.05.2;  
              08.05.3;  
               10.15.1)  
   }
   \title{On the evolutionary status of Be stars}

   \author{Juan Fabregat\inst{1} \and J. Miguel Torrej\'on\inst{2}}
 
   \institute{Departamento de Astronom\'{\i}a, 
Universidad de Valencia, 46100 Burjassot, Valencia, Spain
   \and Departamento de F\'{\i}sica, Ingenier\'{\i}a de Sistemas y
Teor\'{\i}a de la Se\~{n}al, Universidad de Alicante, Spain}

   \offprints{J. Fabregat}
   \mail{juan@pleione.uv.es}
 
   \date{Received date; accepted date}
 
   \maketitle

   \begin{abstract}

We present a study of the abundance of Be stars in open clusters as a
function of the cluster age, using whenever possible ages determined
through Str\"omgren $uvby$ photometry. For the first time in studies of
this kind we have considered separately classical and Herbig Be stars. 

The main results can be summarized as follows:

\begin{itemize}

\item Clusters associated to emitting nebulosities and undergoing stellar
formation are rich in emission line objects, which most likely are all
pre main-sequence stars. No bona fide classical Be star has yet been
identified among them.

\item Clusters younger than 10 Myr and without associated nebulosity are
almost completely lacking Be stars, despite they have a complete
unevolved B main sequence.

\item Classical Be stars appear at an age of 10 Myr, and reach the maximum
abundance in the age interval 14-25 Myr.

\end{itemize}

We interpret our results in the sense that the Be phenomenon is an
evolutionary effect which appears in the second half of the main sequence
lifetime of a B star. We propose that it can be related to main
structural changes happening at this evolutionary phase, which also lead
to the recently discovered non-monotonic helium abundance enhancement.
The semiconvection or turbulent difussion responsible of the surface
helium enrichment, coupled with the high rotational velocity, can generate
magnetic fields via the dynamo effect and thereby originate
the Be phenomenon. Observational tests to this hypothesis are proposed.

   \end{abstract}

\section{Introduction}

The evolutionary status of classical Be stars is a frequently raised and
yet unsolved question. The main issue is to determine whether the Be
phenomenon appears at a given stage in the evolutionary track of every B
star, or it originates in the conditions of formation of some stars, which
include fast rotation and probably other facts. A fundamental element in
this discussion is the study of Be stars in open clusters, in two
different ways: i./ the determination of the Be star positions in the
cluster photometric diagrams; and, ii./ the study of the abundance of Be
stars as a function of the cluster age.

It is well known that Be stars usually occupy anomalous positions in the
colour-magnitude diagrams, lying above the main sequence.
Early attempts to explain the Be phenomenon suggested that Be stars occur
during the core contraction phase following the exhaustion of hydrogen
(Schmidt-Kaler 1964). Later, however, it was observed a significant
fraction of Be stars close to the ZAMS (Schild \& Romanishin 1976), and
today it is generaly accepted that they occupy the whole main sequence
band and different evolutionary states (Mermilliod 1982; Slettebak 1985)
and therefore they are not confined to any particular evolutionary phase.
It is well established that the anomalous positions in the
photometric diagrams can be explained in terms of the contribution of the
circumstellar continuum emission to the photometric indices (Fabregat et
al. 1996; Fabregat \& Torrej\'on 1998; and references therein).  

Extensive studies of the abundances of Be stars in open clusters have been
done by Mermilliod (1982) and recently by Grebel (1997). Both authors
obtained similar results, finding Be stars in clusters of all ages, with a
peak frequency in clusters with turn-off at spectral types B1-B2, and
a regular decrease with increasing age afterwards.

Nevertheless, these kind of studies face some difficulties which make
their conclusions somewhat uncertain. The purpose of this paper is to
critically review the previous work in this field, and to present a new
study of the abundance of Be stars in clusters of different ages taking
into account new considerations and observational results.

\section{Critical review of previous work}

In this section we will address the main drawbacks which affect the
determination of the Be star abundace as a function of the cluster age.
Whenever possible we will propose solutions to these problems.

\subsection{The ages of the open clusters}
The usual way to determine the age of a star cluster is by means of
isochrone fitting. The most extended technique is to transform the
theoretical isochrone from the $L$ - $T_{\rm eff}$ plane to the
observational colour-magnitude plane, and then directly compare it with
the observational photometric data.

In young open clusters the isochrone fitting is difficult by two main
problems affecting the observational data. The usual presence of
differential reddening across the cluster face widens the observed main
sequence. For the clusters we are dealing with in this paper, the presence
of Be stars which generally occupy anomalous positions in the
colour-magnitude diagrams, also contributes to a further main-sequence
widening. Hence, the fit of a particular isochrone can be a very uncertain
process. For instance, recent age determinations for the cluster with the
higest Be star abundance in the Galaxy, NGC 663, are the following: 21 Myr
(Leisawitz 1988), 9 Myr (Tapia et al. 1991), 12-15 Myr (Phelp \& Janes
1994) and 23 Myr (this work, Sect. 3). The difference in the age
determinations amounts to a factor of three.

As an attempt to solve this problem, we have investigated the different
photometric indices which are commonly used as horizontal axis in the
observational HR diagrams, with regard to the B star region of the main
sequence. In Table 1 we present for each index its variation along the B
star sequence, the photometric accuracy which is usually reached in the
photometric data, the sampling of the main sequence -- the ratio between
the index variation and the accuracy -- and how the interstellar reddening
affects the index.

\begin{table}
\begin{center}
\caption{Photometric indices commonly used as $T_{\rm eff}$ indicators in
the observational HR diagrams.} 
\begin{tabular}{ccccc}
\hline
&&&&\\
Index & B range & accuracy & sampling & $E(i)/E(B-V)$ \\
&&&&\\
\hline
&&&&\\
$(B-V)$ & 0.30 & 0.010 & 30 & 1.0 \\
$(U-B)$ & 1.00 & 0.020 & 50 & 0.7 \\
$(V-R)$ & 0.15 & 0.010 & 15 & 0.8 \\
$(V-I)$ & 0.40 & 0.010 & 40 & 1.6 \\
$(b-y)$ & 0.10 & 0.005 & 20 & 0.7 \\
\cuno   & 1.10 & 0.020 & 55 & 0.2 \\
&&&&\\
\hline
\end{tabular}
\end{center}
\end{table} 

In view of this table, the best sampling is obtained with the $(U-B)$
colour in the Johnson system and the \cuno index in the Str\"omgren
system. \cuno has the additional advantage of being much less affected by
reddening. Furthermore, the \mv - \cuno diagram allows an efficient
segregation of the Be stars from the absorption line B stars (Fabregat et
al. 1996). Therefore we propose that the most efficient way to determine
reliable ages for very young clusters is the isochrone fitting to the
observational \mv - \cuno HR diagram. In Section 3 we will base our
discussion of the Be stars abundances as a function of the cluster age on
ages determined in this way.

\subsection{Determination of Be star frecuencies}

Spectroscopic surveys devoted to the detection of Be stars in open
clusters are scanty in the literature. The only clusters for which the
abundance of Be stars have been exhaustively studied are h and $\chi$
Persei (NGC 869 and NGC 884) and NGC 663. The last systematic, although
not exhaustive, survey dates from 1976 with the work of Schild \&
Romanishin.

In the recent years new detection techniques based on CCD
imaging photometry are being  applied to study the Be star abundances in
clusters in the Galaxy (Capilla \& Fabregat 1999) and in the Magellanic
Clouds (Grebel et al. 1992; Grebel 1997; Keller et
al. 1999). These studies, however, only provide lower limits of the
abundance for no more than 10 clusters. We are still far from having a
statistically significant sample of open clusters with well determined Be
star abundances.

\subsection{Classical versus Herbig Be stars}

In this paper we deal only with the abundances of classical Be stars.
There exists other classes of early-type emission line stars. Among them
the most conspicuous are the so called Herbig Ae/Be stars. These objects
are pre-main sequence stars in which the line emission originates in
circumstellar material remaining from the proto-stellar cloud from which
the star was formed (for a recent review of Herbig Ae/Be stars, see
Waters \& Waelkens 1998). The observational
characteristics of classical and Herbig Be stars, at least in the optical
region, are very similar, making a very difficult task to differenciate
between the two types. An efficient segregation can be made in the
far-infrared region, where the Herbig Ae/Be stars show an
important excess caused by the presence of dust, which is lacking in
classical Be stars.

Grebel (1997) includes in her study the clusters NGC 2244, NGC 6611 and IC
2944. All of them are very young open clusters (age $<$ 6 Myr)
associated to bright emission nebulosities, and are still
undergoing stellar formation (Hillenbrand et al. 1993; P\'erez et
al. 1987; Reipurt et al. 1997; de Winter et al. 1997). Hillenbrand et al.
(1993) found a large number (27) of \ha emission line stars in NGC 6611.
They pay special atention to the study of the strong emiters W235 and
W503, and show evidence that these stars are Herbig Ae/Be stars instead of
classical Be stars. For the rest of the stars the situation is much more
uncertain, but they concluded that all emission line stars in NGC 6611 are
pre-main sequence objects instead of classical Be stars. De Winter et al.
(1997), in a smaller sample in the same cluster, found 11 emission line
stars. They classified three of them as Herbig Ae/Be stars, and agreed
with Hillenbrand et al. that most probably the rest of the emission line
objects are also pre-main sequence stars.

In the same way, Reipurth et al. (1997) analysed 7 emission line stars in
the \ion{H}{ii} region IC 2944. They classified all of them as young
pre-main sequence objects, probably Herbig Ae/Be stars. Van den Anker et
al. (1997) found 5 emission line stars in the star-forming cluster NGC
6530. They classified 3 as Herbig Ae/Be, and concluded that the remaining
two most probably are of the same type.

We can conclude that, despite the difficulty of differenciating 
between classical and Herbig Be stars, there are in the literature several
positive identifications of Herbig Ae/Be stars in the youngest, 
star-forming open clusters. Conversely, no bona fide classical Be star
has yet been reported among them.    

\section{The Be star abundance as a function of the cluster age}

There are only four galactic open clusters with a known distinctly high
abundance of Be stars, namely more than 15 Be stars, or more than 25\% of
Be stars among their observed B stars. They are NGC 663, NGC 869, NGC 884
and NGC 3760. As we argued in Sect. 2.1, we will assume for these
clusters the ages determined through isochrone fitting to the
observational \mv - \cuno HR diagram. These ages are 14 Myr for NGC 869
and NGC 884 (Fabregat et al. 1996) and 24 Myr for
NGC 3766 (Shobbrook 1985, 1987). For NGC 663, in Fabregat et al. (1996) we
assumed the age of 21 Myr, because it was the age determination found in
the literature which shown the best agreement with our \uvby data.
However, in Fig. 2 of this reference it can be seen that the 21 Myr
isochrone does not make the best fit to the data in the \mv - \cuno
diagram.
A much better fit is obtained with an age of 23 Myr, and we will assume
this value as the cluster age. We conclude that, when using the isochrone
fitting to the \mv - \cuno diagram as age estimator, the galactic
clusters with the highest frecuency of Be stars occupy the very narrow age
interval of 14-24 Myr.

Grebel (1997) report on three more clusters in the Magellanic Clouds,
with Be stars abundances comparable or even higher than the 
clusters referred to in the above paragraph. The ages of these clusters
fall in the same range than those of the above galactic clusters. They are
NGC 330 (19 Myr), NGC 2004 (20 Myr) and NGC 1818 (25 Myr). Notice that
the age reported by Grebel for NGC 1818 is between 25 and 30 Myr, but
looking at her Fig. 1 we find that the 25 Myr fits better the
data. The same conclusion is reached by Van Veber and Vanbeveren (1997). 

Dieball and Grebel (1998) studied three more clusters in the LMC, namely
SL 538 (18$\pm$2 Myr), NGC 2006 (22.5$\pm$2.5 Myr) and KMHK 19 (16
Myr). They found between 5\% to 12\% of Be stars among the observed
clusters B stars. 

Keller at al. (1999) also searched for Be stars in the clusters NGC 330
and NGC 346 in the SMC, and NGC 1818, NGC 1948, NGC 2004 and NGC 2100 in
the LMC. They found a large amount of Be stars in all these clusters, with
frecuencies ranging from 13\% to 34\% of the clusters main sequence B
stars. For the three clusters not in common with Grebel (1997), the age
determinations in the literature are 15 Myr for NGC 1948
(Vallenari et al. 1993) and 18 Myr for NGC 2100 (Cassatella et al.
1996). We will exclude NGC 346 from this discussion because it is a much
younger cluster, embedded in N66, the largest and brightest \ion{H}{ii}
region in the SMC (Kudritzki et al. 1989; Massey et al. 1989).

All the age determinations for the Magellanic Clouds clusters are derived
from $BVRI$ photometry, and so we consider it less reliable that the ages
obtained from \uvby photometry, for the reasons explained in Sect. 2.3.
Keller et al. consider these ages uncertain by factors of 2 to 3. Despite
of this fact, the coincidence between these ages and the age interval we
determined from the \uvby photometry of the galactic clusters is
overwhelming. Only one Magellanic Clouds cluster fall very marginally
outside of the age interval of 14-24 Myr, namely NGC 1818 (25 Myr). 

In Table 2 we resume the ages and Be star abundances of
all discussed "Be star rich" clusters. When comparing the abundances of
the Galactic and Magellanic Clouds clusters it should be noted that
abundances in the latter are derived from single-epoch surveys, while in
the former they came from more than 50 years of study. In fact, it turns
out that the Be star abundance of the Magellanic Clouds clusters is
significantly higher, and this can be attributed to the
different metalicity (Maeder et al. 1999).

\begin{table}
\begin{center}
\caption{The clusters with the highest Be star abundance in the Galaxy and
Magellanic Clouds.} 
\begin{tabular}{lrrl}
\hline
  & & & \\
Cluster & Age & N$_{\rm Be}$/N$_{\rm OB}$ & Ref. \\ 
  & & & \\ 
\hline
  & & & \\
Milky Way &&& \\
  & & & \\
NGC 663 & 23 & 40\% & Sanduleak (1990) \\
NGC 869 & 14 & 25--50\% & Waelkens et al. (1990) \\
NGC 884 & 14 & 25--50\% & Waelkens et al. (1990) \\
NGC 3766 & 24 & 36\% & Shobbrook (1985, 1987) \\
  & & & \\
SMC &&& \\
  & & & \\
NGC 330 & 19 & 34\% & Keller et al. (1999) \\
  & & & \\
LMC &&& \\
  & & & \\
NGC 1818 & 25 & 21\% & Keller et al. (1999) \\
NGC 1948 & 15 & 14\% & Keller et al. (1999) \\
NGC 2004 & 20 & 13\% & Keller et al. (1999) \\
NGC 2006 & 22 & 12\% & Dieball \& Grebel (1998) \\
NGC 2100 & 18 & 28\% & Keller et al. (1999) \\
SLC 538  & 18 & 12\% & Dieball \& Grebel (1998) \\
KMHK 1019 & 16 & 5\% & Dieball \& Grebel (1998) \\
  & & & \\
\hline
\end{tabular}
\end{center}
\end{table}

From the above, we conclude that the clusters with a high abundance of Be
stars occupy a very narrow range of ages, namely between 14 and 25 Myr.
For older clusters the percentage of Be stars decrease, as shown by
Mermilliod (1982) and Grebel (1997). We have to study now the abundances
in the younger clusters. There are not in the literature enough
observational data to make a complete analysis, but we have been able to
collect several pieces of evidence pointing towards the same conclusion:
the very paucity of Be stars in clusters younger than 10 Myr. In the
following paragraphs we will review some of them.

We have performed a first search in the WEBDA database of open cluster
data (Mermilliod 1999). We found 64 clusters younger than 10 Myr. Among
them, 9 clusters contain 3 or more Be stars. They include  NGC 2244 (3 Be
stars), NGC 6530 (18), NGC 6611 (20) and IC 2944 (8). All these clusters
have been discussed in Sect. 2.3, where we show that their emission line
objects are Herbig Be stars instead of classical Be stars. NGC 6823 (5)
lies in a bright \ion{H}{ii} region with associated dark clouds (Stone
1988). NGC 7380 (3) is associated to the molecular cloud regions Sh2-142
and NGC 7380 E, and contains pre-main sequece stars among which several
Herbig Ae/Be stars are identified (Chavarr\'\i a-K. et al. 1994). IC
1590 (4) is embedded in the nebulosity of NGC 281, also identified as the
bright \ion{H}{ii} emission region Sharpless 184 (Guetter \& Turner 1997).
One more cluster, IC 1805 (2) is located in an \ion{H}{ii} region, and
associated to a large molecular cloud (Ninkov et al. 1995; Heyer et al.
1996). For the same arguments exposed in Sect. 2.3, we consider that the
emission line objects in the last four clusters are likely to be pre-main
sequence objects. The two remaining Be star rich clusters are NGC 884 (17)
and NGC 957 (5). NGC 884 has been discussed in the above paragraphs, where
we assumed and justified an age of 14 Myr. Meynet et al.
(1993) also derived an age of 14 Myr from isochrone fitting to $UBV$
photometric data. The age of 8 Myr in the WEBDA
database is clearly wrong. The same can be said for NGC 957, whose age in
the database is 6 Myr. This cluster has an age of 15 Myr in the last
edition of the Lyng{\aa } catalogue (Lyng{\aa } 1987), which would place
it in the age range we have established for the maximum Be star abundance.
Maeder, Grebel \& Mermilliod (1999) assumed ages of 13 and 16 Myr
respectively for NGC 884 and NGC 957. 

Among the remaining 54 clusters younger than 10 Myr in the WEBDA database,
one contain two Be stars (NGC 6871) and three one Be star (NGC 6383, NGC
7235 and Hogg 16). Even in these very few cases doubts still remain
on the cluster ages and the nature of the emission line objects.
The emission line star in NGC 6383 has been studied by Th\'e et al.
(1985), who cannot decide whether it is a classical Be star or a pre-main
sequence object. A recent determination of the age of Hogg 16 gives
a result of 25 Myr (V\'azquez \& Feinstein 1991). 50 more clusters
younger than 10 Myr in the WEBDA database have no Be stars detected so
far.

Balona and co-workers (Balona 1994; Balona \& Koen 1994; Balona \& Laney
1995, 1996) obtained CCD \uvbyb photometry for several young open
clusters. They did not make any particular investigation of the Be star
content of the clusters they observed, but we have searched their
photometric lists for objects with emission in the \hb line. 

A $\beta$ index equal to 2.55 correspond to an equivalent width of the \hb
line equal to 0. i.e., a photospheric absorption line completely filled-in
by  emission (Fabregat \& Torrej\'on 1999). Hence, $\beta < 2.55$
indicates that the \hb line is in emission. We have searched the above
referred to photometric lists for OB stars (\byo $<$ 0.05) with $\beta <
2.55$.
This search will also detect, as well as classical Be stars, other kinds
of early-type emission line objects, like Of and OBIa stars. To exclude
these stars we have introduced the additional restriction of $M_V > -4.5$.
A proof of the reliability of this last restriction
can be found in the following data: the two emission line stars in NGC
3293 brighter than $-$4.5 have spectral classification in the literature;
they are star 3, with type B0.5Ib (Feast 1958), and star 4, type B0Ib
(Morgan et al. 1955). In NGC 6231 there are 8 emission line stars brighter
than $-$4.5, and  7 among them have spectral types given by Levato \&
Malaroda (1980). Two are OBI supergiants and three more Of. Even if the
remaining three are Be stars -- Oe in this case -- this would not affect
the main conclussions of this work, as we will comment on later.

60\% of the Be stars in NGC 663, NGC 869 and NGC 884 observed by Fabregat
et al. (1996) would have been detected by applying the above criteria.
This percentage can be considered as the typical detection capability of a
photometric survey. Photometric surveys never detect the whole content of
Be stars by the two following reasons: i./ surveys through photometric
filters only detect stars with high level of line emission, loosing the
mild emitters which only can be identified by spectroscopic means; ii./
the Be phenomenon is variable, and at a given time only a fraction of Be
stars are in a phase of line emission.

After all these considerations, the final results of our survey in the
photometric data published by Balona and co-workers are given in Table 4. 
Except in the case of NGC 2362, the ages reported have been derived from
the pulsational properties of the $\beta$ Cephei stars present in each
cluster by Balona et al. (1997).
As it can be seen the clusters in the age interval 4-10 Myr are almost
lacking of Be stars. No Be stars have been found in the two younger
clusters, with ages of 4-5 Myr. A few have been detected in the two
older, with ages of 9-12 Myr.

\begin{table}
\begin{center}
\caption{Clusters with CCD \uvbyb photometry obtained by Balona and
co-workers. For each cluster we indicate the age, number of OB stars
observed and number of Be stars.} 
\begin{tabular}{cccc}
\hline
  & & \\
Cluster & age & N$_{\rm OB}$  & N$_{\rm Be}$ \\ 
  & & & \\ 
\hline
  & & & \\
NGC 2362 & 5$\pm$2 & 33 & 0 \\
NGC 3293 & 9.1$\pm$0.2 & 136 & 2 \\
NGC 4775 & 11.7$\pm$1.5 & 121 & 3 \\
NGC 6231 & 3.7$\pm$0.6 & 129 & 0 \\
 & & & \\
\hline
\end{tabular}
\end{center}
\end{table}

\section{Discussion}

The general picture which emerges from the analysis in the previous
section is the following: star forming clusters, associated to bright
emission nebulae, are rich in emission line stars, but they are much
likely pre-main sequence objects related to Herbig Ae/Be stars. When the
nebula dissipates the process of star formation stops -- at least with
regard to the massive stars -- and the clusters are devoid of
early-type emission line objects. Classical Be stars start to appear in
clusters with age of around 10 Myr, and reach their maximum abundance in
the 14-25 Myr interval. For older clusters the Be star abundance decreases
with age, as shown by Mermilliod (1982) and Grebel (1997).

The decreasing of the Be star abundance with the age after the 14-24 Myr
peak is a reflect of the dependence of the Be star abundance with the
spectral type. It is well known that the maximum abundance occurs for
spectral type B1-B2 (Zorec \& Briot 1997). Clusters older than 25 Myr have
their turnoff at type B3 or later, and hence they are expected to contain
lower abundances than clusters with B1-B2 main sequence stars. Clusters
older than 100 Myr have their turnoff at B8 or later, and the lack of Be
stars is an obvious reflect of the lack of any kind of B stars.

Conversely, the lack of Be stars in clusters younger than 10 Myr has
evident implications on the evolutionary status discussion. These clusters
have their turnoff at type B1 or earlier, and hence they have their B star
sequence complete, including the spectral types for which the Be star
abundance reach its maximum. The lack of Be stars in these clusters
implies that a Be star cannot be a very young object. 

Be stars appear in clusters with turnoff at B1, and reach its maximum
abundance in clusters with turnoff at B2. As most of the Be stars belong
to these types, we have to conclude that Be stars are much closer to the
end of the main sequence than to the ZAMS.  

This result contradicts the finding of Mermilliod (1982) and Slettebak
(1985), already mentioned in the introduction, who stated that Be stars
occupy the whole main sequence band from the ZAMS to the TAMS. This
afirmation is mainly based on photometric data, in the $UBV$ system, of Be
stars in open clusters. For a B star of a given subtype, the difference in
$(B-V)$ between its position at the ZAMS and the end of the main sequence
is lower than 0.1 mag. To firmly conclude that a Be star is in or near the
ZAMS, a photometric accuracy of 0.02 mag. for the underlying star of the 
Be object would be required. If we consider all the problems which affect
photometry of Be stars, this accuracy seems not to be within reach.
Mermilliod (1982) use the $(U-B)$ colour, for which the diference between
ZAMS and TAMS is higher. But Be stars tend to move leftwards in the
\mv - $(U-B)$ diagram due to the excess in the $U$ magnitude caused by the
circumstellar emission in the Balmer continuum (Kaiser 1989). The same
effect is present in the Str\"omgren \cuno index, as shown by Fabregat et
al. (1996). This effect can displace a strong emitter from TAMS to ZAMS
and even leftwards. Mermilliod already realized this effect when he states
that the $(U-B)$ colours are affected by the Be phenomenon.
Hence we consider our result based on the
analysis of Be star abundances in open clusters more reliable that
the results based on photometric data which are strongly affected by the
circumstellar continuum emission. On the other hand, both authors are
aware that most Be stars occur on the evolved part of the main sequence
(Mermilliod 1992) and considerably off the ZAMS (Slettebak 1985).

There is additional evidence indicating that Be stars are somewhat evolved
objects. In the younger clusters containing early-type Be stars, late type
Be stars are scarce or completely lacking. This was first noted by
Sanduleak (1979, 1990). He found 26 Be stars in NGC 663, and among them
only 2 later than B5. His objetive prism survey was complete to mag. 14,
which at the cluster reddening and distance reach the spectral type
B6-B7V. He concludes that Be stars in NGC 663 are primarily
confined to spectral types earlier than B5.

Capilla \& Fabregat (1999) performed CCD Balmer-line photometry of NGC
663, NGC 869 and NGC 884. Their images are deep enough to cover all the B
type range. They detected a total of 25 Be stars in the three clusters,
and among them only two are later than B5. 

These results can be interpreted in the same evolutionary terms that
before. In clusters with ages in the interval 14-24 Myr, the stars earlier
than B5 have spent more than a half of its life in the main sequence,
while the late B stars are still in the first half of the main sequence
phase. The Be phenomenon accurs among the former and not among the latter.

\subsection{Be stars as post-mass-transfer binary systems}

It has been suggested that Be stars could be the result of the evolution
of close binary systems. The transfer of matter and angular momentum would
produce the spin up of the mass gainer to very high rotation rates. It is
well known that rapid rotation is a common characteristic of Be stars, and
hence a key ingredient of the Be phenomenon. The products of close binary
evolution are therefore good candidates to develope the Be phenomenon
(Pols et al. 1991). Moreover, several Be stars are definitely
post-mass-transfer systems. They are the Be/X-ray systems, in which a
neutron star orbits an early-type Be star, accreting matter from the dense
stellar wind and thereby generating X-rays. The properties of Be stars in
Be/X-ray binaries are not different from those of the rest of Be stars.
For a recent review of the properties of Be/X-ray binaries, see Negueruela
(1998).

Our conclusions on the evolutionary status of Be stars are consistent
with the hypotesis of the nature of Be stars as post-mass-transfer
systems. The Be phenomenon would occur after the mass transfer phase in
the evolution of a close binary. This would explain the scarcity of Be
stars in very young clusters: the necessary time for the main sequence
evolution of the primary star in the system has to be over before the mass
transfer begins and the Be star is formed, and hence a Be star cannot be
a very young object.

However, the interpretation of the Be phenomenon as the result of close
binary evolution faces important problems, both theoretical and
observational. The computations of Pols et al. (1991) only can account for
about half the population of Be stars. The recent study of Van Bever \&
Vanbeveren (1997) with updated models of close binary evolution reveal
that only a minority of the Be stars (less than 20\% and possibly as low
as 5\%) can be due to close binary evolution. On observational grounds,
the models of close binary evolution predict a population of Be star plus
white dwarf systems ten times more abundant than the Be/X-ray binaries.
These systems should be observable as low luminosity X-ray sources. The
search conducted by Meurs et al. (1992) failed in detecting the predicted
population of Be+WD systems.

Hence we have to conclude that, despite the consistency with the results
of our analysis, the model of the close binary evolution does not provide
a satisfactory explanation to the Be star phenomenon. Moreover, it has to
be considered that this model is ad hoc, because it only justifies the
formation of a rapidly rotating B star, but does not explain how the Be
phenomenon arises from it.

\subsection{Evolution through the main sequence}

The main conclusion of our study is that Be stars are evolved main
sequence stars, closer to the TAMS than to the
ZAMS. This would imply the existence of some evolutionary change able to
produce the Be star phenomenon during the main sequence lifetime.
This is not in agreement with the classical theory of stellar evolution,
which predicts that the main sequence is a quiet evolutionary stage in
which no major changes in the stellar structure occur. 

However, in the modern literature there is a growing evidence of
important changes which occur during the main sequence stage. Lyubimkov
(1996,
1998) has shown that the abundance of helium and nitrogen in O and early B
stars increases during the main sequence. This change is not monotonic.
The initial helium abundance $He/H$ = 0.08-0.09 is maintained during the
first half of the main sequence lifetime. Subsequently, $He/H$ abruptly
increases approximately twofold in a short interval of relative ages \ra
(where $t_{\rm MS}$ is the main sequence lifetime) between 0.5 and 0.7,
and this enhanced $He/H$ remains constant until the main sequence stage is
complete. Recent evolutionary models take into account this effect, and
attribute the light element enhancement as due to early mixing 
produced by rotationally induced turbulent diffusion (Denissenkov 1994; 
Talon at al. 1997; Maeder 1997).

Our results are consistent with the Be phenomenon appearing at the same
evolutive age \ra $\sim$ 0.5. To show this we have to keep in mind that
the highest percentage of Be stars corresponds to spectral types B1-B2
(Zorec \& Briot 1997). 
The age of 10 Myr, where Be stars start to appear, correspond to an
evolutionary age of \ra = 0.5 for a star of about 10 \sm and solar
metalicity (this and next  paragraph discussion is based on data
from Table 47 and Fig. 1 in Schaller et al. 1992). Such a star
at this
age has an spectral type of B1, where the abundance of Be stars reach its
maximum. In clusters younger than 8 Myr, stars are
reaching the TAMS at 20 \sm or higher, and at spectral types earlier than
B1. The lower abundance of Be stars among these earlier types, and the
low relative number of such massive stars explain the paucity of Be stars
among these very young  clusters. O8-9.5e stars can, however, start to
appear at as early an age as 3 Myr, which can explain the few
cases of clusters younger than 10 Myr with a few Be stars. NGC 6231,
discussed in Section 3, could be one of this cases.

The maximum Be star abundance occurs between 14 and 25 Myr. 14 Myr
correspond with \ra = 0.6 for a 9 \sm star, at spectral type B1, i.e. at
the begining of the maximum abundance of Be stars. 25 Myr is the end of
the main sequence lifetime for a 9 \sm star, at spectral type B3. Stars of
lower mass reach the relative age of \ra = 0.5 at spectral types of B3 or
later, where the abundance of Be stars decreases. This explains the
decreasing abundance of Be stars after an age of 25 Myr.

We propose the hypothesis that the Be phenomenon is an evolutionary
effect, appearing half way of the main sequence
lifetime, and is related to the light elements enhancement which occurs at
the same evolutionary phase. It is now widely accepted that magnetic
fields near the stellar surface could be the cause of the enhanced mass 
loss which characterize the Be phenomenon. The mechanisms proposed to
explain the mixing
at this stage, which imply movement of plasma near the stellar surface,
coupled with the rotation of the star, could originate and maintain a
magnetic field via a dynamo related effect. The characteristic high
rotational velocity of Be stars would play a major role in: i./ inducing
the movement of matter via turbulent diffusion; and, ii./ enhancing the
magnetic field strenght via the dynamo effect. Hence our hypothesis
provides a natural explanation of the influence of high rotational
velocity in the Be phenomenon.

A direct proof of this hypothesis could be obtained by studying whether Be
stars have an enhanced helium abundance. Unfortunately, the contamination
of the photospheric spectrum by the circumstellar emission lines makes the
abundance analysis of Be stars an almost impossible task. Such analysis
can be performed in Be stars observed in a disk-loss phase, i.e., in a
phase in which the circumstellar disk has dissipated and the photospheric
spectrum is directly observable. This has been done by Lyubimkov et al.
(1997) for the Be star X Persei, and they obtained an enhanced helium
abundance of $He/H$ = 0.19. However, this case is not conclusive, because
X Persei is a Be/X-ray binary which in the past underwent mass transfer, 
and hence it is not possible to know whether
the helium overabundance is due to internal processes in the current
primary star or whether it is caused by external accretion 
from the original primary. Helium abundance studies of isolated Be stars
observed during disk-loss phases are required to proof our sugestions. 
 
\section{Conclusions}

We have presented a study of the abundance of Be stars in open clusters
as a function of the cluster age, using whenever possible ages determined
through Str\"omgren $uvby$ photometry. For the first time in studies of
this kind we have considered separately classical and Herbig Be stars. 

The main results obtained can be summarized as follows:

\begin{itemize}

\item Clusters associated to emitting nebulosities and undergoing stellar
formation are rich in emission line objects, which most likely are all
pre-main sequence objects. No bona fide classical Be star has yet been
identified among them.

\item Clusters younger than 10 Myr and without associated nebulosity are
almost completely lacking Be stars, despite they have a complete
unevolved B main sequence.

\item Classical Be stars appear at an age of 10 Myr, and reach the maximum
abundance in the age interval 14-25 Myr.

\end{itemize}

We have interpreted our results in the sense that the Be phenomenon is an
evolutionary effect which appears in the second half of the main sequence
lifetime of a B star. This conclusion is supported by other facts, like
the lack of late-type Be stars in young clusters rich in early-type Be
stars.

We propose the hypothesis that the Be phenomenon could be related to main
structural changes happening at an evolutionary age of \ra = 0.5, which
also lead to the recently discovered non-monotonic helium abundance
enhancement. The semiconvection or turbulent difussion responsible of the
helium and nitrogen enrichment, coupled with the high rotational velocity,
can originate magnetic fields via the dynamo effect. It is now widely
accepted that many observed phenomena are due to Be star photospheric
activity related to the presence of magnetic fields. Our hypotesis
provides
a natural explanation of the relationship between the Be phenomenon and
the high rotational velocity characteristic of Be stars.

It should be noted, however, that our results on the Be star
frequencies in open clusters came from scarce and inhomogeneous sets of
data, and this leads to a somewhat speculative component in our
conclusions. To check our results and proposed explanations the following
observational data would be of critical importance:

\begin{itemize}

\item A systematic study of the Be star frequencies in a significant
number of clusters of different ages, both in the Galaxy and in the
Magellanic Clouds. It
would be of exceptional interest to know the abundances in Magellanic
Clouds clusters younger than 10 Myr and with high-mass stellar formation
finished yet.

\item The determination of cluster ages in an homogeneous system. We
propose for this purpose the use of the Str\"omgren $uvby$ system.

\item The determination of the helium and light elements abundance of Be
stars undergoing phases of disk-loss.

\item In a more general context, the determination of the helium
abundance in B stars with evolutionary ages \ra $>$ 0.5. Early type B
stars in clusters with turn-off at B1-B2 would be specially well suited
for this purpose.

\end{itemize}

We are currently undertaking observational programs to address the
above questions.

To conclude, we would like to comment that in the recent years, many
relations and cross-links between classical Be stars and several other
types of peculiar hot stars have been put forward, clearly showing that
the Be phenomenon is not just an isolated problem of the stellar
astrophysics. On the other hand, the referred to results on light element
enhancements in the atmospheres of hot stars during their main sequence
lifetime are not compatible with the classical stellar evolutionary 
models, and demostrate that there are important lackings in our knowledge
of massive star evolution. In this paper we propose that both phenomena
are related, and hence, the understanding of the Be phenomenon could be
the clue for the advance in our understanding of major issues in massive
star formation and evolution.

\begin{acknowledgements}
This research has made use of the SIMBAD database, operated at CDS,
Strasbourg, France, and the NASA's Astrophysics Data System Abstract
Service.
\end{acknowledgements}


\begin{thebibliography}{}
   \bibitem{} van den Ancker M.E., Th\'e P.S., Feinstein A. et al. 1996,
      A\&AS

   \bibitem{} Balona L.A. 1994, MNRAS 267, 1060

   \bibitem{} Balona L.A., Koen C. 1994, MNRAS 267, 1071

   \bibitem{} Balona L.A., Laney C.D. 1995, MNRAS 276, 627

   \bibitem{} Balona L.A., Laney C.D. 1996, MNRAS 281, 1341

   \bibitem{} Balona L.A., Dziembowski W.A., Pamyatnykh A. 1997, MNRAS
      289, 25

   \bibitem{} Capilla G., Fabregat J. 1999, A\&A (submitted)

   \bibitem{} Cassatella A., Barbero J., Brocato E., Castellani V., Geyer
      E.H. 1996, A\&A 306, 125

   \bibitem{} Chavarr\'\i a-K. C., Moreno-Corral M.A., Hern\'andez-Toledo
      H., Terranegra L., de Lara E. 1994, A\&A 283, 963

   \bibitem{} Denissenkov 1994, A\&A 287, 113

   \bibitem{} Fabregat J., Torrej\'{o}n J.M. 1998, A\&A 332, 643 

   \bibitem{} Fabregat J., Torrej\'{o}n J.M. 1999, PASP (submitted) 

   \bibitem{} Fabregat J., Torrej\'{o}n J.M., Reig P., et al. 1996, A\&AS
      119, 271 

   \bibitem{} Feast M.W. 1958, MNRAS 118, 618

   \bibitem{} Grebel E.K. 1997, A\&A 317, 448

   \bibitem{} Grebel E.K., Richtler T., de Boer K.S. 1992, A\&A 254, L5

   \bibitem{} Guetter H.H., Turner D.G., 1997, AJ 113, 2116

   \bibitem{} Heyer M.H., Brunt C., Snell R.L., et al. 1996, ApJ 464, L175 

   \bibitem{} Hillenbrand L.A., Massey P., Strom S.E., Merrill K.M. 1993,
      AJ 106, 1906

   \bibitem{} Kaiser D. 1989, A\&A 222, 187

   \bibitem{} Keller S.C., Wood P.R., Bessell M.S. 1999, A\&AS 134, 489

   \bibitem{} Kudritzki R.P., Cabanne M.L., Husfeld D. et al. 1989, A\&A
      226, 235

   \bibitem{} Leisawitz D. 1988, "Catalog of Open Clusters and Associated
      Interstellar Matter", NASA-RP 1202

   \bibitem{} Levato, H., Malaroda S. 1980, PASP 92, 323

   \bibitem{} Lyng{\aa } G. 1987, Catalogue of Open Cluster Data, 5th
      Edition, Lund Observatory

   \bibitem{} Lyubimkov L.S., 1996, A\&SS 243, 329 

   \bibitem{} Lyubimkov L.S., 1998, Astr. Rep. 42, 52

   \bibitem{} Lyubimkov L.S., Rostopchin S.I., Roche P., Tarasov A.E. 
      1997, MNRAS 286, 549

   \bibitem{} Maeder A., 1997, A\&A 321, 134

   \bibitem{} Maeder A., Grebel E.K., Mermilliod J.C. 1999, A\&A 346, 459

   \bibitem{} Massey P., Parker J.W., Garmany C.D. 1989, AJ 98, 1305

   \bibitem{} Meynet G., Mermilliod J.C., Maeder A. 1993, A\&AS 98, 477 

   \bibitem{} Mermilliod J.C., 1982, A\&A 109, 48

   \bibitem{} Mermilliod J.C., 1999, in: Very Low-Mass Stars and Brown
      Dwarfs in Stellar Clusters and Associations. R. Rebolo and M.R.
      Zapatero-Osorio (eds.), Cambridge Univ. Press.

   \bibitem{} Meurs E.J.A., Piters A.J.M., Pols O.R. et al. 1992, A\&A
      265, L41 

   \bibitem{} Morgan W.W., Code A.D., Withford A.E. 1955, ApJS 2, 41

   \bibitem{} Negueruela I. 1998, A\&A 338, 505

   \bibitem{} Ninkov Z., Bretz D.R., Easton R.L., Shure M. 1995, AJ 110,
      2242 

   \bibitem{} P\'erez M.R., Th\'e P.S., Westerlund B.E. 1987, PASP 99,
      1050

   \bibitem{} Phelps R.L., Janes K.A. 1994, ApJS 90, 31

   \bibitem{} Pols O.R., Cot\'e J., Waters L.B.F.M., Heise J. 1991, A\&A
      241, 419 

   \bibitem{} Reipurth B., Corporon P., Olberg M., Tenorio-Tagle G.,
      1997, A\&A 327, 1185

   \bibitem{} Sanduleak N. 1979, AJ 84, 1319

   \bibitem{} Sanduleak N. 1990, AJ 100, 1239

   \bibitem{} Schaller G., Schaerer D., Meynet G., Maeder A. 1992, A\&AS
      96, 269

   \bibitem{} Schild R.E., Romanishin W. 1976, ApJ 204, 493

   \bibitem{} Schmidt-Kaler T., 1964, Bonn Verr\"off. 70, 1

   \bibitem{} Shobbrook R.R. 1985, MNRAS 212, 591

   \bibitem{} Shobbrook R.R. 1987, MNRAS 225, 999

   \bibitem{} Slettebak A. 1985, ApJS 59, 769

   \bibitem{} Stone R.C. 1988, AJ 96, 1389 

   \bibitem{} Talon S., Zahn J.P., Maeder A., Meynet G. 1997, A\&A 322,
      209

   \bibitem{} Tapia M., Costero R., Echeverr\'\i a J., Roth M. 1991, MNRAS
      253, 649

   \bibitem{} Th\'e P.S., Hageman T., Westerlund B.E., Tjin A Djie H.R.E. 
      1985, A\&A 151, 391

   \bibitem{} Vallenari A., Bomans D.J., de Boer K.S. 1993, A\&A 268, 137

   \bibitem{} Van Bever J., Vanveberen D. 1997, A\&A 322, 116 

   \bibitem{} V\'azquez R.A., Feinstein A. 1991, A\&AS 90, 317

   \bibitem{} Waelkens C., Lampens P., Heynderickx D. et al. 1990, A\&AS
      83, 11 

   \bibitem{} Waters L.B.F.M., Waelkens C. 1998, ARA\&A 36, 233 

   \bibitem{} de Winter D., Koulis C., Th\'e P.S. et al. 1996, A\&AS 

   \bibitem{} Zorec J., Briot D. 1997, A\&A 318, 443

\end{thebibliography}
\end{document}